# How Magnetic Field Inhomogeneity of Plane Wiggler Affects Spontaneous Radiation of Electrons

E.A. Ayryan [1], A.H. Gevorgyan [2], M. Hnatic [3,4,5],
N.Sh. Izmailian [6], V.V. Papoyan [3], K.B. Oganesyan [1,6*]

[1] Joint Institute for Nuclear Research, LIT, Dubna, Moscow Region, Russia
[2] Far Eastern Federal University, Russky Island, Vladivostok, Russia
[3] BLTP, Joint Institute for Nuclear Research, Dubna, Russia
[4] Faculty of Sciences, P. J. Safarik University, Kosice, Slovakia
[5] Institute of Experimental Physics SAS, Kosice, Slovakia
[6] Yerevan Physics Institute, Alikhanyan National Science Lab, Yerevan, Armenia

*bsk@yerphi.am

The spectral distribution of spontaneous emission of electrons moving in a plane wiggler with inhomogeneous magnetic field is calculated. We show that electrons do complicated motion consisting of slow(strophotron) and fast(undulator) parts. The equations of motion are averaged over fast undulator part and we obtain equations for connected motion. It is shown, that the account of inhomogenity of the magnetic field leads to appearance of additional peaks in the spectral distribution of spontaneous radiation.

## 1. INTRODUCTION

The free electron lasers (FEL) are high power tunable sources of coherent radiation that are used in science research, for heating of plasma, in the physics of condensed media, in atomic, molecular and optical physics, in biophysics, biochemistry, biomedical engineering etc. The radiation produced by present-day FELs has a range from the millimeter to the X-rays waves, which no other similar high intensity tunable sources cover [1, 2]. In modern science this branch is of interest both from the viewpoint of fundamental research and application.

In FELs [3, 4] the kinetic energy of relativistic electrons moving through the spatially modulated magnetic field of a wiggler is used for production of coherent radiation. The frequency of radiation is determined by the energy of electrons, the space period and the magnetic field strength of a wiggler. This permits a retuning of FEL to be made in a wide range as opposed to the atomic and molecular lasers. In a conventional FEL the magnetic field of wiggler is constant, but it is inhomogeneous in the transverse direction [5]. It is important to take into account this inhomogeneity that causes a complicated motion of electrons: fast undulator oscillations along the wiggler axis and slow strophotron ones [6–14] in the transverse direction.

In the present work we describe the equations of motion of electrons moving along the wiggler axis, the magnetic field in which is spacially inhomogeneous. The aim of the work is to calculate the spectral distribution of spontaneous radiation.

## 2. EQUATIONS OF MOTION

The vector potential of the magnetic field of undulator has the form [15]

$$\mathbf{A}_w = -\frac{H_0}{q_0}\cosh q_0 x \sin q_0 z \mathbf{j}, \tag{1}$$

where $H_0$ is the magnetic field strength, $q_0 = 2\pi/\lambda_0$, $\lambda_0$ the wiggler period, $\mathbf{j}$ the unit vector in $y$ direction. Bellow we consider the problem in the paraxial approximation

$$q_0 x \ll 1. \tag{2}$$

Taking into account (2) the expression for magnetic field (1) acquires the form:

$$H_x = H_0\left(1+\frac{q_0{}^2 x^2}{2}\right)\cos q_0 z; \quad H_y = 0; \quad H_z = H_0 q_0 x \sin q_0 z. \tag{3}$$

The equations of motion ($c = 1$)

$$d\mathbf{p}/dt = e[\mathbf{vH}] \tag{4}$$

in magnetic field (3) have the form:

$$\ddot{x} = -\frac{eH_0 q_0}{\varepsilon} x\dot{y}\sin q_0 z,$$

$$\ddot{y} = \frac{eH_0}{\varepsilon}\left[\dot{z}\left(1+\frac{q_0{}^2 x^2}{2}\right)\cos q_0 z + q_0 x\dot{x}\sin q_0 z\right], \tag{5}$$

$$\ddot{z} = \frac{eH_0}{\varepsilon}\left(1+\frac{q_0{}^2 x^2}{2}\right)\dot{y}\cos q_0 z,$$

and the energy change is

$$d\varepsilon/dt = 0, \quad \varepsilon = \text{const}.$$

Expression (5) was obtained taking into account that $p_{x,y,z} = v_{x,y,z}\varepsilon$.

It is seen that

$$\left(\frac{q_0 x^2}{2}\sin q_0 z\right)' = q_0 x\dot{x}\sin q_0 z + \frac{q_0{}^2 x^2}{2}\dot{z}\cos q_0 z, \tag{6}$$

$$\int \dot{z}\cos q_0 z dt = \int \cos q_0 z dz = \frac{\sin q_0 z}{q_0}. \tag{7}$$

Using relations (6) and (7) we obtain after integration of the second equation of (5)

$$\dot{y} = \frac{eH_0}{\varepsilon q_0}\left(1 + \frac{q_0{}^2 x^2}{2}\right)\sin q_0 z\,. \tag{8}$$

By substitution of (8) in the first and third equations of (5) we obtain with due regard for (2)

$$\begin{cases} \ddot{x} = -\left(\dfrac{eH}{\varepsilon}\right)^2 x \sin^2 q_0 z, \\ \ddot{z} = -\dfrac{1}{2q_0}\left(\dfrac{eH}{\varepsilon}\right)^2 \sin 2q_0 z\left(1 + q_0{}^2 x^2\right). \end{cases} \tag{9}$$

Averaging the first equation of (9) with respect to the wiggler period $2\pi / q_0$ and taking into account that $\overline{\left(\sin^2 q_0 z\right)} = 1/2$, we have

$$\ddot{x} + \Omega^2 x = 0\,. \tag{10}$$

The latter has a solution

$$x = a_0 \cos\left(\Omega t + \theta_0\right), \tag{11}$$

where

$$\Omega = \frac{eH_0}{\sqrt{2\varepsilon}}, \quad a_0 = \sqrt{x_0{}^2 + \frac{\alpha^2}{\Omega^2}}, \quad \cos\theta_0 = \frac{x_0}{a_0}, \quad \sin\theta_0 = -\frac{\alpha / \Omega}{a_0}. \tag{12}$$

The averaging of the second equation of (9) gives

$$\left(\ddot{z}\right)^{(0)} = 0, \quad \left(\dot{z}\right)^{(0)} = v, \quad \left(z\right)^{(0)} = vt. \tag{13}$$

Taking into account (11) and (13) the second equation of (9) admits a solution

$$\delta z = -\frac{\Omega^2}{2q_0{}^2}t + \frac{\Omega^2}{2q_0{}^3}\sin 2q_0 t + \frac{a_0{}^2\Omega^2}{16q_0}\sin\left\{2\left(q_0 + \Omega\right)t + 2\theta_0\right\} + \frac{a_0{}^2\Omega^2}{16q_0}\sin\left\{2\left(q_0 - \Omega\right)t - 2\theta_0\right\}. \tag{14}$$

Thus, for $z = z^{(0)} + \delta z$ we have

$$z = t\left(1 - \frac{1}{2\gamma^2} - \frac{\Omega^2}{2q_0{}^2}\right) + \frac{\Omega^2}{4q_0{}^3}\sin 2q_0 t + \frac{a_0{}^2\Omega^2}{16q_0}\sin\left\{2\left(q_0 + \Omega\right)t + 2\theta_0\right\} + \\ + \frac{a_0{}^2\Omega^2}{16q_0}\sin\left\{2\left(q_0 - \Omega\right)t - 2\theta_0\right\}. \tag{15}$$

Here the allowance was made for the fact that $1 - v = 1/\left(2\gamma^2\right)$, where $\gamma = \varepsilon/\left(mc^2\right)$ is the relativistic factor, $m$ the electron mass, $c$ the velocity of light and $\varepsilon$ the electron energy.

The obtained results are valid in case of the following approximations:

$$a_0 q_0 < 1, \quad \frac{\Omega}{q_0} < 1, \quad a_0\Omega < 1. \tag{16}$$

The electrons perform fast (undulator) oscillations in the longitudinal direction (along the wiggler axis), whereas in the transverse one they perform slow (strophotronic) oscillations in the direction of $x$. axis and fast (undulator) oscillations in the direction of $y$ axis.

## 3. SPONTANEOUS RADIATION

Using the solutions for $x(11)$, $\dot{y}(8)$ and $z(15)$ one can find the spectral intensity of spontaneous radiation, that in the direction of $z$ axis (wiggler axis) is determined by formula [16]

$$\frac{d\varepsilon}{d\omega do} = \frac{e^2\omega^2}{4\pi^2}\left|\int_0^T dt\left[\mathbf{n}\times\mathbf{v}\right]e^{i\omega(t-z)}\right|^2, \tag{17}$$

where $do$ is an infinitesimal solid angle in the direction of $z$ axis and $T$ is the time of electron flight through the undulator.

Using formula [17]

$$e^{-iA\sin x} = \sum_{n=-\infty}^{\infty} J_n(\mathrm{A})e^{-inx} \tag{18}$$

with Bessel functions $J_n(\mathrm{A})$, and omitting cumbersome calculations we obtain

$$\frac{d\varepsilon}{d\omega do} = \frac{e^2\omega^2\Omega^2T^2}{8\pi^2 q_0{}^2}\sum_{n,m,k=-\infty}^{\infty}\left(I_{n+1,k,m} - I_{n,k,m}\right)^2\left(\frac{\sin u}{u}\right)^2, \tag{19}$$

where

$$\left[\begin{array}{l} u = \dfrac{T}{2}\left[\omega\left(\dfrac{1}{2\gamma^2}+\dfrac{\Omega^2}{2q_0{}^2}\right)-(2\mathrm{n}+1)q_0 - 2m\Omega\right], \\ I_{n,k,m} = J_{n-k}\left(Z_1\right)J_{\frac{k+m}{2}}\left(Z_2\right)J_{\frac{k-m}{2}}\left(Z_2\right), \\ Z_1 = \dfrac{\omega\Omega^2}{4q_0{}^3}, Z_2 = \dfrac{\omega a_0{}^2\Omega^2}{4q_0}. \end{array}\right. \tag{20}$$

Equation (19) describes the radiation spectrum consisting of a superposition of spectral lines localized at combinated frequencies of odd harmonics $(2\mathrm{n}+1)\omega_{\mathrm{res,und}}$ of undulator resonant frequency and even harmonics $2m\omega_{\mathrm{res,str}}$ of strophotronic resonant frequency, where $m,n = 0,1,2,3,\ldots$. Here

$$\omega_{\mathrm{res,und}} = \frac{2\gamma^2 q_0}{1+\gamma^2\,\Omega^2/q_0{}^2}, \quad \omega_{\mathrm{res,str}} = \frac{2\gamma^2\Omega}{1+\gamma^2\,\Omega^2/q_0{}^2}. \tag{21}$$

## 4. CONCLUSION

It is shown that as a result of the allowance for magnetic field inhomogeneity, some additional peaks appear in the spectral distribution of spontaneous radiation. Out of the multitude of these peaks one can obtain ultrashort pulses using the well-known mode-locking method. The peaks in the spectral distribution of spontaneous radiation are localized at combinated frequencies of the odd harmonics of $(2n+1)\omega_{res,und}$ undulator resonant frequency and even harmonics of $2m\omega_{res,str}$ strophotron resonance frequency. In case of an undulator with constant magnetic field the peaks are localized at odd harmonics of undulator resonant frequency, and in case of a strophotron they are localized at odd harmonics of strophotronic resonant frequency. One may conclude thus, that due to the presence of inhomogeneity in the magnetic field in the plane wiggler these two systems (wiggler with the constant magnetic field and the strophotron) are integrated in one unit and there appear peaks in the spectral distribution of spontaneous radiation at combined (odd undulator and even strophotronic) resonant frequencies.